\documentclass[conference]{IEEEtran}

\usepackage{epsfig}
\usepackage{epstopdf}
\usepackage{graphicx}
\usepackage{url}
\usepackage{changebar}
\usepackage{xcolor}
\usepackage{soul}

\newif\ifdiff
\difftrue % remove this line to apply the changes without highlighting them

\ifdiff
\newcommand{\removed}[1]{\cbstart\removedfragile{#1}\cbend{}}
\newcommand{\removedfragile}[1]{{\color{red}\st{#1}}{}}
\else
\newcommand{\removed}[1]{} % non-markup version

\fi

\begin{document}

\title{DART-MPI: An MPI-based Implementation of a PGAS Runtime System}

\author{
  \IEEEauthorblockN{
    Huan Zhou\IEEEauthorrefmark{1},
    Yousri Mhedheb\IEEEauthorrefmark{2},
    Kamran Idrees\IEEEauthorrefmark{1},
    Colin W. Glass\IEEEauthorrefmark{1},
    Jos\'e Gracia\IEEEauthorrefmark{1},
    Karl F{\"u}rlinger\IEEEauthorrefmark{3} 
    and Jie Tao\IEEEauthorrefmark{2}
  }
  \IEEEauthorblockA{
    \IEEEauthorrefmark{1}High Performance Computing Center Stuttgart (HLRS), University of Stuttgart, Germany
  }
% {\it \{huan.zhou, idrees, gracia\}@hlrs.de} \\
  \IEEEauthorblockA{
    \IEEEauthorrefmark{2}Steinbuch Center for Computing, Karlsruhe Institute of Technology, Germany
  }
%{\it \{yousri.mhedheb, jie.tao\}@kit.edu} \\
  \IEEEauthorblockA{
    \IEEEauthorrefmark{3}Department of Computer Science, Ludwig-Maximilians-Universit{\"a}t (LMU) M{\"u}nchen, Germany
  }
%{\it fuerling@nm.ifi.lmu.de} \\
}

\maketitle

\begin{abstract}

A Partitioned Global Address Space (PGAS) approach treats a
distributed system as if the memory were shared on a global
level. Given such a global view on memory, the user may program
applications very much like shared memory systems.  This greatly
simplifies the tasks of developing parallel applications, because no
explicit communication has to be specified in the program for data
exchange between different computing nodes. In this paper we present DART, a
runtime environment, which implements the PGAS paradigm on large-scale
high-performance computing clusters.  A specific feature of our
implementation is the use of one-sided communication of the Message
Passing Interface (MPI) version 3 (i.e. MPI-3) as the underlying
communication substrate. We evaluated the performance of the
implementation with several low-level kernels in order to determine
overheads and limitations in comparison to the underlying MPI-3.

\end{abstract}


\section{Introduction}

The traditional way to parallelize a program for a distributed memory
system is to use an explicit approach to coordinate the data
distribution and movement. I.e., the programmer has to assign data to
the processes and organize the data movement across the computing
nodes using primitives provided by the programming model. This is not
an easy task, especially for today's applications with large amounts
of data, complicated data structures, and the stringent requirements
on optimization of communication patterns in order to achieve
scalability on large machines.

Conceptually, developing a parallel program is simpler in a shared
memory programming model.  Rather than using explicit send/receive
pairs to exchange data, processes communicate which each other
implicitly through shared memory. For instance, a producer can write
data into shared memory, while a consumer accesses the data with a
read operation in much the same way as the data is accessed in
a sequential program, however the programmer needs to use certain synchronization mechanism, 
such as lock, semaphore or monitor, in order to 
guarantee the conflict-free accesses to the shared data. 
%This asserts the correctness of the shared data when multiple threads are modifying it.
Besides simplifying the parallelization process,
this allows for automatic tuning of the data assignments (and
associated performance) at compile time.

The basis for shared memory programming is formed by a global memory
space, which is not physically available on clusters and other
distributed memory architectures. To enable shared memory style
programming on such machines, an early approach was to build a virtual
or distributed shared memory layer on top of the distributed
memory~\cite {hrunting98,santo2010,ramesh1997}. Therefore, the Partitioned Global Address Space (PGAS) model~\cite
{pgas}, was proposed and is gaining popularity.

The PGAS model provides an abstraction of a global memory address space, that is logically partitioned. 
Each portion of the global address space has an affinity to a certain process or thread.
%PGAS addresses data locality awareness with a goal of better
%performance achieved by reducing the data movement between the nodes.
A number of PGAS programming systems have been implemented in the
past, including Unified Parallel C (UPC)~\cite {upc}, Co-Array Fortran
(CAF)~\cite {oriCAF,newCAF}, Chapel~\cite {chapel}, X10~\cite {X10Spec}, STAPL~\cite{STAPL} and Titanium~\cite
{titanium}. Typically, these approaches rely on one-sided
communication substrates such as GASNet~\cite {techgasnet} to perform
inter-node communication. GASNet is a low-level networking library
which implements remote-memory access primitives and is thus
particularly suitable for PGAS models.

In the context of the DASH project~\cite{dash}, we have developed a
runtime system interface for supporting shared memory style programming on
distributed memory systems called DART (the DASH runtime). DASH is a
C++ template library to efficiently work with distributed data
structures. DASH implements PGAS semantics through operator
overloading and supports the allocation of and parallelization over
large data sets and provides means of achieving multi-level
hierarchical data locality~\cite{Fuerlinger:2014:DASH}.

DART provides the C++ library DASH with services and abstracts from a
variety of underlying communication substrates. For our scalable DART
runtime implementation we chose MPI, more specifically MPI-3, as the
underlying communication mechanism and we call this implementation
DART-MPI while DART-SHMEM and DART-CUDA are other implementations currently
under development at the authors' organizations.

We chose MPI because it is a standard, well-developed communication
substrate, with support for different network technologies. In
general, MPI implementations, in particular those provided by system
vendors, are highly optimized for their particular network fabric.
MPI introduced the concept of one-sided communication, called RMA
(Remote Memory Access), in the second version of its
specification. The RMA features were improved in the third version
(MPI-3). Our PGAS runtime benefits from the optimized implementations
of MPI-3 on different architectures as well as its support for
one-sided non-blocking inter-node communications.

In this paper we discuss the semantic gap between  MPI-3 RMA 
and DART, describe the implementation details of the runtime system
DART-MPI, evaluate the performance and scalability of DART-MPI on
a Cray XE6 supercomputer using a number of low-level benchmarks and finally
discuss our approach in the context of similar works. 
%We validated the DART concept and measured the communication latency,
%bandwidth and synchronization overhead.

The remainder of the paper is organized as follows: Section \ref
{related} gives a brief overview of related work, followed by the API
specification of our runtime system in Section \ref {api}. Section
\ref {implementation} describes the implementation using MPI-3. In
Section \ref {evaluation} benchmark results are shown and
discussed. Finally, the paper concludes with a short summary and
future work in Section \ref {conclusion}.

\section {Related Work}
\label {related}

%The PGAS approach establishes a global memory space on a system with
%memory distributed across the computing nodes. For the programmers
%the memory address covers the entire system, however, the data for
%computation may reside on remote memory. This means that the data
%must be transferred via the interconnecting network. PGAS languages
%rely on a communication substrate to perform the inter-node
%communications.

The most popular PGAS languages are UPC~\cite {upc}, CAF~\cite {oriCAF,newCAF},
Chapel~\cite {chapel}, X10~\cite {X10Spec}, STAPL~\cite{STAPL} and Titanium~\cite {titanium}. UPC is one of the
first and one of the few fully implemented PGAS languages 
and is an extension of the C programming language. CAF 1.0~\cite {oriCAF} is
an extension to Fortan 95 for SPMD parallel processing. It coverts
Fortran 95 into a robust parallel language with a few rules related to
two fundamental issues: work distribution and data
distribution. By 2005, the  Fortran Standards Committee decided to make CAF 2.0~\cite {newCAF}
integrated into the Fortran 2008 standard. Compared to the CAF 1.0, CAF 2.0 can present a richer set of 
coarray-based language extensions. Chapel (Cascade High-Productivity Language) is a product
of Cray Inc., developed as part of the DARPA High Productivity
Computing Systems (HPCS) program. Chapel is not an extension of
existing languages but a stand alone block-structured
language. 
X10 is designed as an object-oriented parallel programming language, 
through which the Asynchronous PGAS(APGAS)~\cite{APGAS} is realized. X10 need to leverage a runtime system 
that named X10RT~\cite{X10RT}  for doing the underlying communications.
X10RT is presented as a C library and can be implemented in different forms,
in which the X10RT-MPI is realized on top of MPI-2. 
The STAPL (Standard Template Adaptive Parallel Library)
is a productive framework for C++, 
it provides support for developing parallel program on both shared and 
distributed memory system.
Titanium is an explicitly parallel dialect of Java that
extends Java by immutable classes, multidimensional arrays, an
explicitly parallel SPMD model of computation with a global address
space, and zone-based memory management.

Besides PGAS languages there are also approaches that implement PGAS
in the form of an API and a library. An example is SHMEM, a library
API that allows its participating processes to view a partitioned
global address space. It was started by Cray Inc.\ in 1993 and adopted
by other vendors later. Currently, the OpenSHMEM community project
\cite {openshmem} is building a new and open specification to
consolidate the various existing SHMEM versions into a widely accepted
standard. Global Arrays (GA) \cite {garray} has originally been
developed over 20 years ago and provides one-sided global data access
for regularly structured one- or multi-dimensional arrays.

Many of the PGAS languages mentioned above adopt GASNet~\cite {techgasnet} as one of the
options for the underlying communication library. The reference
implementation of OpenSHMEM is also based on GASNet. GA uses ARMCI
(Aggregate Remote Memory Copy Interface) \cite{armci} as its primary
communication layer.

GASNet (Global Address Space Networking) is a language-independent,
low-level networking layer that provides network-independent,
high-performance communication primitives. It is tailored for
implementing a parallel global address space and is therefore not
surprisingly the most common choice.

The Message Passing Interface (MPI) is a standardized and
portable message passing library, based on the consensus of the MPI
Forum \cite{mpi} organized by vendors, library developers, researchers and
users. MPI has been widely and commonly used for parallel programs on
HPC platforms. However, MPI is most commonly used for two-sided
communication that involves both the sender and the receiver of a
message. PGAS languages or libraries require direct remote memory
access (RMA) to shorten the access latency to remote memories. Hence,
two-sided communication is not efficient to meet the characteristics of RMA 
in PGAS languages (see for instance \cite{Bonachea}).

RMA, which was added with MPI-2, has introduced the basic concept of
\emph{windows} to specify a local memory region accessible to remote
processes, enabling one-sided communication. These MPI-2 RMA
operations have, however, been found to be too limiting and lacking
for adoption in PGAS programming systems. Bonachea et
al.~\cite{Bonachea} describe the reasons why the traditional MPI-1
two-sided primitives as well as the extended RMA interfaces introduced
in MPI-2 are insufficient for PGAS models. In addition, they list a
set of useful and constructive suggestions for improving the MPI-2 RMA
semantics.

For studying the potential of using MPI for PGAS an Integrated Native
Communication Runtime supporting both MPI and UPC communication on
Infiniband Clusters is proposed by Jose et al.~\cite{Jose}. It is
observed that the integrated runtime is capable to rival the existing
UPC runtime based on GASNet. However, all the experiments conducted in
this study identified a common limitation that is based on the
Infiniband architecture. Additionally, the pitfalls of portability to
PGAS models, including UPC, stayed unchanged.

Daily et al.~\cite{Daily} explored four alternative methods to check
the suitability of using MPI-2 in implementing the RMA communications,
including put, get and atomic memory operations, in PGAS models. They
found that the two-sided semantics require an implicit synchronization
between sender and receiver. Additionally the strict limitation on
suboptimal implementation of MPI-2 RMA leads to a severe degradation
of performance.

Dinan et al.~\cite{James} developed techniques for overcoming the
semantic mismatches between MPI-2 RMA and ARMCI, and presented a
complete implementation of an ARMCI runtime system on MPI-2
RMA. However, the benchmarks demonstrated that MPI-2 RMA failed to
gain any obvious advantages over ARMCI in performance.

%But the MPI-2 one-sided communication API shows inadequacy for
%implementing PGAS \cite {dan}.  

%For example, the window creation in MPI-2 is a collective operation,
%where all processes that intend to use a window (for RMA) must
%participate, making it unfeasible to have a window for each shared
%object.  there are no alternative for programmers but only to create
%MPI window by using the call of {\em MPI\_Win\_Create}, which will
%reduce the flexibility and efficiency potentially.  For instance,
%concurrent conflicting RMA access to the same location and concurrent
%access to overlapping windows are not allowed in MPI-2. In addition,
%%RMA operations to a given window are only permitted to access the
%memory of a single process.  we only can start a passive RMA access
%epoch to a single process rather than all the processes in the given
%window. Overall, the MPI-2 specification provides one-sided
%communication for RMA, but is still not mature for implementing PGAS.

To address the limitations identified for MPI-2, an extended and
revised set of RMA operations was defined for MPI-3~\cite{Balaji,
  mpi3}. In MPI-3, any allocated memory is private to the MPI process
and can be exposed to other processes as a public memory region. Two
new window allocation functions are introduced: a collective version
to allocate windows for fast access and a dynamic version which
exposes no memory but allows the user to register remotely accessible
memory locally and dynamically at each process. Two memory models are
available to allow the implementation to benefit from
cache-coherency. In addition, MPI-3 provides mechanisms for
performance optimization, such as atomic operations.

A recent publication~\cite{mpi3shmem} studies an OpenSHMEM
implementation based on MPI-3, focusing mostly on mapping the OpenSHMEM
one-sided interfaces to the MPI-3 ones \cite{mpi3shmem}. The
micro-benchmarks show that OSHMPI performs better than MVAPICH2-X in
\emph{get} latency on a shared memory system (intra-node), there is
however still room for improving OSHMPI in the case of \emph{put}
operations and distributed memory (inter-node).
%which is however lack of designing related to balance with the
%semantics of MPI-3 other than RMA, such as team and group. %TODO: I
%do not understand the previous sentence at all (Colin)
  
%\todo{We have named a number of reasons not to use MPI3 for PGAS, so
%  why did we?}  We decided to use MPI-3 as the communication library
%  to implement DART. Recentl%y, research has been published
%  concerning the challenges and potential problem%s with using MPI as
%  communication system for PGAS models and on resulting desig%n
%  opportunities.  Our runtime DART aims at bridging the semantic gap
%  between the MPI-3 RMA and th%e DASH design without lowering the
%  performance. The following two sections firs%t give an overview of
%  the DART interface and then the implementation details.

\section {The DART Application Programming Interface}
\label {api}

DART is a plain C based interface on which the C++ template library
DASH is built. DART provides services to the DASH library, defines
common concepts and terminology and abstracts from the underlying
communication substrate and hardware. While DART is hidden from users
of the high level DASH library, it can also be used directly by users
or form the basis for other PGAS projects. Therefore, we use the term
API (Application Programming Interface) for our interface. An overview
of DART is presented in the following. The complete DART specification
is available online at \url{http://www.dash-project.org/dart}.

The main task for DART is to establish a partitioned global address
space and to provide functions to handle memory efficiently, such as
memory allocation and data movement. In addition, DART also provides
functions for initialization, synchronization and management of
teams. The DART API is divided into the following five parts:

\begin{itemize}
\item Initialization and shutdown
\item Team and group management
\item Synchronization
\item Global memory management
\item Communication
\end{itemize}

For initialization and shutdown DART provides the functions {\em
  dart\_init} and {\em dart\_exit}. In addition, functions for
querying the environment are also contained in this part of the
interface specification.

DART provides interfaces to support team and group management. In a
DASH/DART program the individual participants are called
\emph{units}. Each unit has a non-negative zero-based integer ID that
remains unchanged throughout the program execution. A DART unit is
similar to an MPI process or a UPC thread. We use the generic term
``unit'' to underline the possibility of mapping a unit to an
OS process, a thread or any other concept that may fit. 

%DART teams and DART groups are formed from multiple units.

A DART \emph{team} is an ordered set of units, identified by an
integer ID. In each application there is a default team that contains
all units comprising the program. A team can have sub-teams and a unit
can belong to several teams and sub-teams. The sub-teams IDs have to
be unique with regard to their parent team ID.
%Note, that contrary to MPI ranks, DASH units have the same ID in all
%teams, i.e. the member units of a team are not numbered contiguously
%within the team. -- this is not true, the local id of a unit is in a
%team is always 0,...,n-1 - Karl

A DART \emph{group} is also an ordered set of units. The difference
between groups and teams is that groups have local meaning only, while
teams are coherent across several units. In other words, group-related
operations are local, while operations to manipulate teams are
collective (and potentially more expensive). DART groups are
essentially helper objects representing sets of units out of which
teams can be formed. Therefore, DART groups function similarly as MPI
groups. %A difference is that DART requires the units in a group to be
%arranged in an ascending order based on their absolute unit IDs.

The DART team/group part of the specification contains common
functions for creating, destroying and querying teams and
groups. These functions are: {\em dart\_group\_init}, {\em
  dart\_team\_create}, {\em dart\_team\_get\_group}, {\em
  dart\_team\_myid} and {\em dart\_team\_size}. In addition, there is
a set of group-related functions for merging or splitting groups, and
modifying the membership.

DART provides functions for synchronization. In addition to collective
synchronization functions like {\em dart\_barrier}, DART also provides
functions for managing mutexes, in order to synchronize shared memory
writing and reading among the DART units.

Providing and working with a global memory is the focus of the
runtime. DART uses several terms to identify memory spaces and the
data located on them. The local address space of a unit is managed by
the regular OS mechanisms and data items are addressed by regular
pointers. The global address space is a virtual abstraction, with each
unit contributing a part of its local memory. Data items are addressed
by global pointers provided by DART. The DART global pointers are
presented with 128 bits, consisting of a 32 bit unit ID, a 16 bit
segmentation ID, 16 bit flags and a 64 bit virtual address or offset.

The terms private and shared describe the accessibility of data items
in DART, where a shared datum can be accessed by multiple units and a
private datum is visible only to one unit.  The terms non-collective
and collective are introduced to differentiate two kinds of DART
global memory allocations.  DART provides a set of functions for
memory allocation in the global address. As the typical DART
non-collective global memory allocation call --- {\em dart\_memalloc}
only allocates a memory region with specified size in the global
address space of the calling unit and returns a \textit{non-collective
  global pointer} to it. As the typical DART collective global memory
allocation call --- {\em dart\_team\_memalloc\_aligned} is a
collective function within the specified team. Each team member calls
the function to request an amount of memory, which is only accessible
to those team members.  The return value of this function is a
\textit{collective global pointer}, pointing to the beginning of the
allocation. There are also functions for freeing the memory and
setting the global pointers.  The terms aligned and symmetric are used
to describe DART collective global memory allocations.  A collective
global memory allocation is called symmetric when the same amount of
memory is allocated by each member of the team and is expected to be
aligned when the same offset can be used in a global pointer to refer
to any member's portion of the allocated memory. A collective global
memory allocation with the characteristics of aligned and symmetric
has the advantageous property that any member of the team can locally
compute a global pointer to any location in the allocated memory.

%The DART communication class consists of functions that perform
%one-sided communications. These include blocking operations like {\em
%dart\_get\_blocking} and {\em dart\_put\_blocking} as well as
%non-blocking operations called {\em dart\_get} and {\em
%dart\_put}. For the latter, DART provides functions, i.e., {\em
%dart\_wait/waitall} and {\em dart\_test/testall}, to check if the
%transfer is completed. DART has also collective operations for data
%exchange within a team, such as {\em dart\_gather}, {\em
%dart\_scatter} and {\em dart\_bcast}.

The DART communication functions consist of one-sided communications
and collective communications.  On the one hand, DART one-sided
communications include blocking operations like {\em
  dart\_get\_blocking} and {\em dart\_put\_blocking} as well as
non-blocking operations called {\em dart\_get} and {\em
  dart\_put}. The DART blocking operations do not return until the
data transfers complete both at the origin locally and at the target
remotely. In addition, for the DART non-blocking operations, DART
provides functions, i.e., {\em dart\_wait/waitall} and {\em
  dart\_test/testall}, to check whether the message transfers are
completed before the data items are applied. On the other hand, DART
collective communications are provided for data exchange within a
team, for example, {\em dart\_gather}, {\em dart\_scatter}, {\em
  dart\_bcast} and so on.

\section {Implementation with MPI-3}
\label {implementation}

We begin with an overview of the MPI-3 standard, and then depict the way of applying MPI-3 to the DART implementation. At the end we examine step-by-step the challenges of balancing MPI-3 and DART in semantics, devise methods of overcoming those challenges and describe the detailed development of implementing DART on the MPI-3 RMA basis.

%\subsection{The Message Passing Interface}
MPI has become the de-facto communication standard for parallel programming, and it is believed to be so popular due to its capability of delivering acceptable and portable performance for diverse underlying network topologies. 

%\todo{Most of this paragraph is variations from material presented in
%Related work (and not underfed by references). Merge with intro or
%related work, in particular the reference at the end of the paragraf.}
%In 1990, the second version of the MPI standard (MPI-2) added support for remote memory access (RMA) operations on top of the two-sided and collective communication models. However, the MPI-2 RMA model functioned in an inflexible fashion in order to follow the feature of portability embodied in the MPI standard. 
%In addition, the conservative memory model defined in MPI-2 constrained the usability of hardware abilities to a great extent, such as the automatic cache coherence on a coherent memory subsystem. Several additional serious limitations related to the MPI-2 RMA access prohibit MPI-2 RMA to be applied efficiently as the underlying communication subsystems of higher-level programming models, such as PGAS. In 2013, the MPI-3 standard was introduced. It included a significant extension to the RMA model, based on one-sided communication. This extension greatly enhanced the usability and effectiveness of the MPI RMA. Therefore, MPI-3 RMA is suitable to be utilized as a runtime system for a variety of PGAS models \cite{Balaji}.

\subsection{Extensions of RMA Model in MPI-3}
MPI \textit{window} is a critical concept for the MPI RMA communication operations. 
The window encompasses a memory region that is exposed to all MPI processes in its associated communicator. 
The typical MPI \textit{window} creation operation --- {\em MPI\_Win\_create} proceeds in a collective way. 
Each process in the given communicator generates a window in its own memory and returns a window object. Besides {\em MPI\_Win\_create}, 
MPI-3 provides three other MPI window creation operations, namely {\em MPI\_Win\_allocate}, {\em MPI\_Win\_allocate\_shared} and {\em MPI\_Win\_create\_dynamic} respectively, 
to generate more specific or flexible MPI windows.

MPI-3 not only supports three basic non-blocking RMA communication
calls --- {\em MPI\_Put}, {\em MPI\_Get} and {\em MPI\_Accumulate},
but also provides counterparts that return request handles, i.e.  {\em
  MPI\_Rput}, {\em MPI\_Rget} and {\em MPI\_Raccumulate}. MPI-3 RMA
supports two kinds of synchronization modes -- active and passive
target. The passive mode does not require the target to participate
explicitly in synchronization operations. Hence, the passive mode is
closer in semantics to an asynchronous communication model than the active mode.
DART utilizes the passive synchronization mode.

\begin{figure}[!ht]
\begin{center}
\includegraphics[scale=0.55]{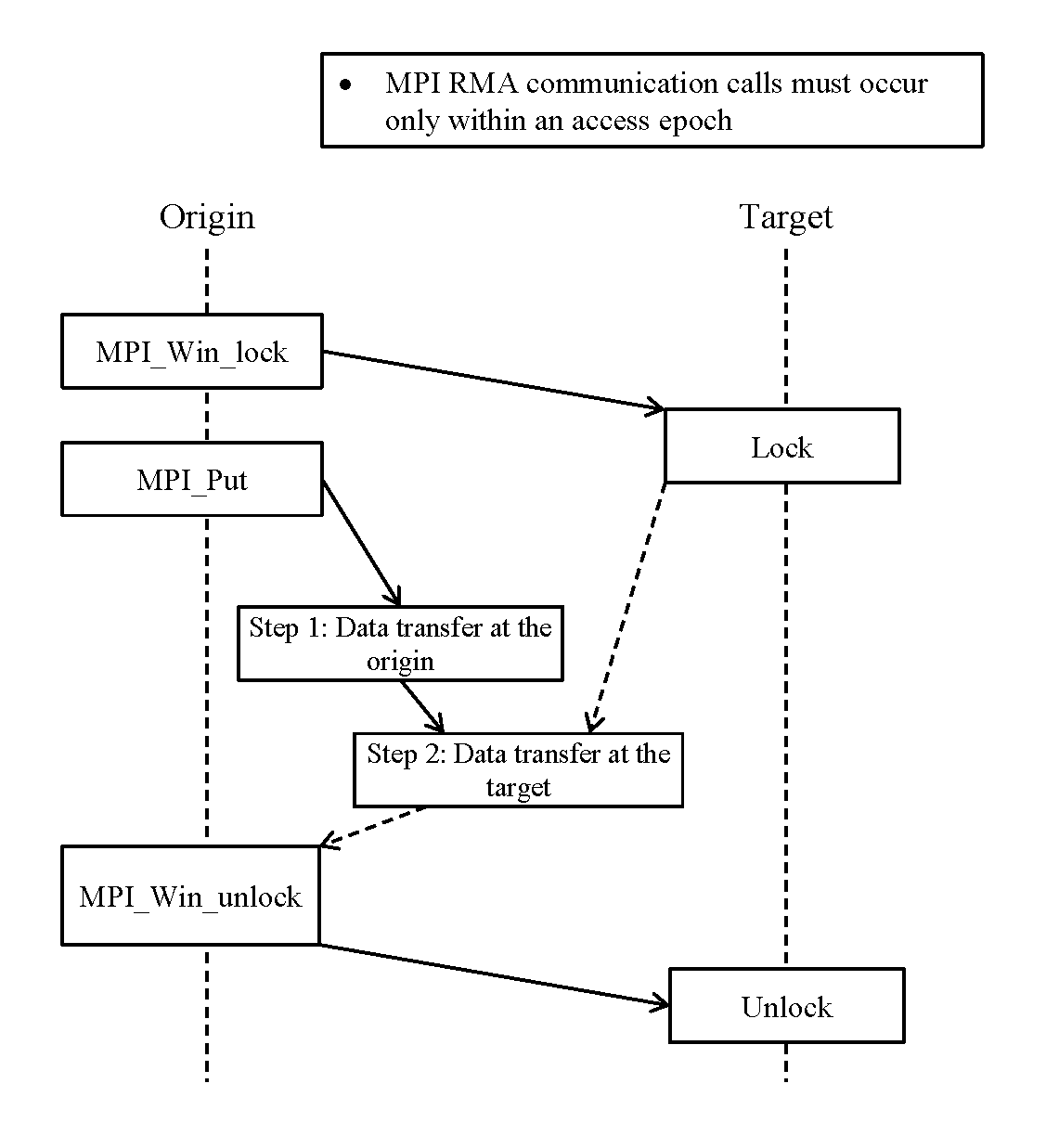} 
\caption{Synchronization Events of MPI Passive Target}
\label{sync}
\end{center}
\end{figure}

As illustrated in Fig. \ref {sync}, the MPI passive mode occurs
within an access epoch which should be initiated by locking the RMA
window and terminated by unlocking it again. Furthermore, passive mode
supports two kinds of lock modes --- shared and exclusive. Exclusive
lock prevents concurrent accesses from distinct processes even for
non-overlapping memory locations in the target window and thus
impairs the concurrency of RMA operations. To maximize concurrent memory access, shared lock is the better choice. %TODO: according to this description - why would anyone use exclusive??? (Colin)
However, the MPI-2 restrictive RMA shared lock semantics greatly limits the behaviors of a RMA passive target with shared lock. 
The following two common accessing operations are forbidden$\colon$ a) two distinct remote operations concurrently updating the same location in a target window; 
b) a remote operation and a local operation concurrently accessing the memory encompassed by a target window. 
Compared with such restrictive semantics, MPI-3 allows the above two cases to happen without any errors but rather an undefined outcome.

Two kinds of window copies -- public and private -- are utilized to address the concept of MPI-3 RMA. For this, two memory models are formed, i.e., RMA unified and RMA separated, according to whether these two copies are be visible to each other or not. The MPI-2 RMA semantics follows a strict rule that only supports the RMA separated model, where public and private copies are required to be always synchronized explicitly (not visible to each other) even on hardware with a coherent memory system. This limitation is removed in MPI-3 with the RMA unified memory model. In the unified memory model, public and private copies can be maintained consistent automatically (visible to each other), which fully matches with the semantics of our runtime DART and potentially improves performance significantly.

\subsection{DART with MPI as the Runtime Substrate}
It appears that MPI-3 RMA matches DART perfectly with its relatively relaxed, flexible and portable semantics. However, there are still several but non-trivial semantic gaps between them, which have to be bridged in an effective way. 

\subsubsection{Create and Sort Group}
\begin{figure}[!ht]
\begin{center}
\includegraphics[scale=0.55]{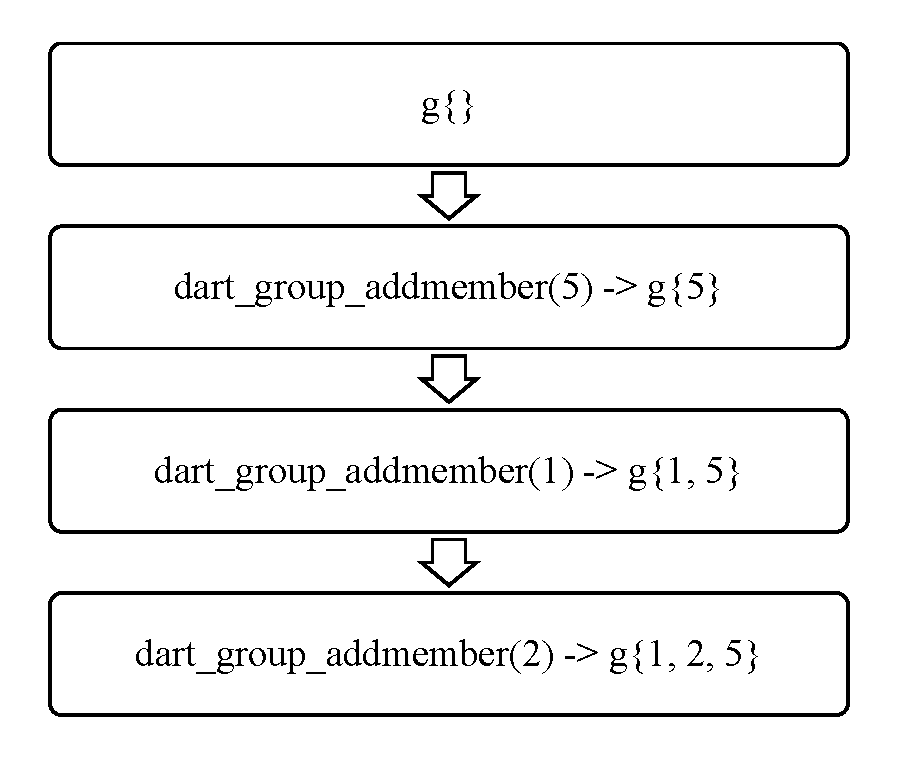}
\caption{Schematic Example of DART Group Creation}
\label{group}
\end{center}
\end{figure}

DART only supports a non-collective mode of group creation --- {\em dart\_group\_addmember}. From Fig. \ref{group} it can be observed that in any case DART group creations are performed on absolute \textit{unitID}s. Additionally, DART groups must be sorted and maintained in an ascending order based on the absolute \textit{unitID}. %TODO: Check this - I vaguely remember having read the opposite above (Colin)

\begin{figure}[!ht]
\begin{center}
\includegraphics[scale=0.55]{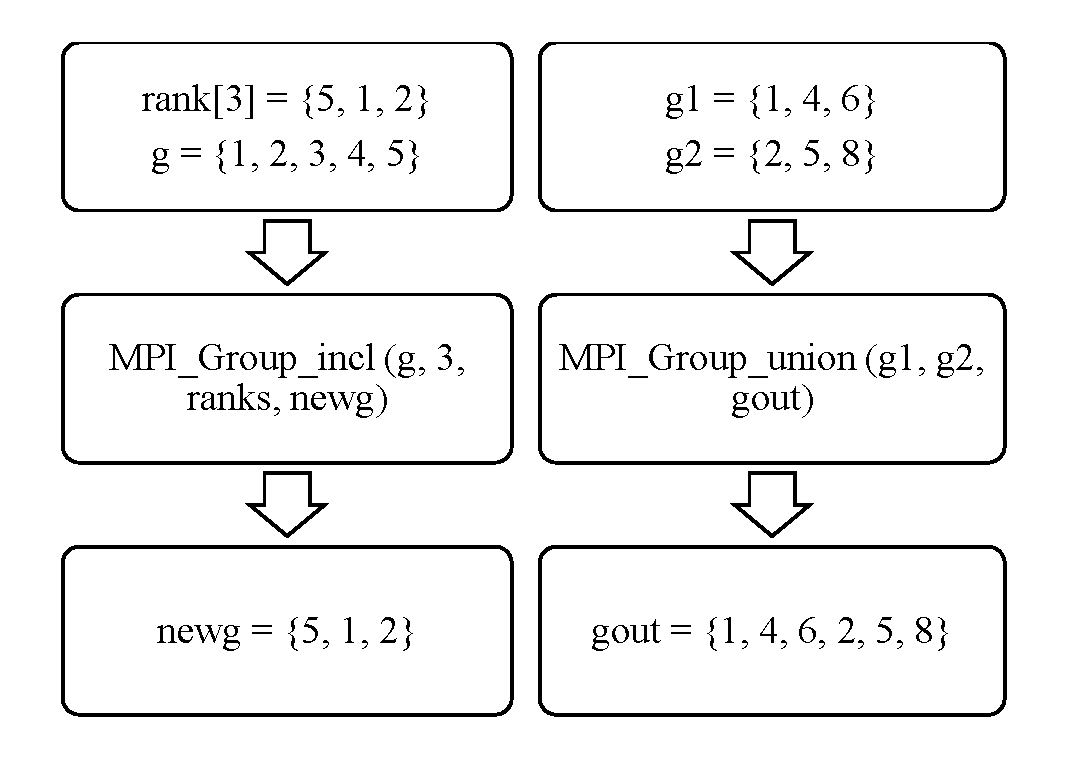}
\caption{Schematic Example of MPI Group Creation}
\label{mpigroup}
\end{center}
\end{figure}

Contrarily, an MPI group is created via a collective function --- {\em MPI\_Group\_incl (parentgroup, n, ranks, newgroup)}. $newgroup$ is comprised of the $n$ elements specified by the array $ranks$ indicating $n$ processes with relative IDs --- $ranks[1],..., ranks[n-1]$ in the parent group. Therefore, the process with rank $i$ in $newgroup$ is the process with rank $ranks[i]$ in the parent group. The MPI group creation mechanism implies two facts that do not fit into the DART group concept. First, a sub-group is created based on the relative ranks in the parent group rather than the absolute ranks (in {\em MPI\_COMM\_WORLD}). Second, the ordering of the processes in a sub-group depends on the ordering in $ranks$. Furthermore, the MPI group union mechanism {\em MPI\_Group\_union (g1, g2, gout)} simply appends $g2$ onto $g1$ instead of guaranteeing the ordering of processes in the output group $gout$. We can conclude that for all practical purposes, the processes in each MPI group are arranged in a random 
fashion. Fig. \ref{mpigroup} illustrates how MPI group creation and group union work.

As a result of the above differences between DART and MPI in handling groups, it is not feasible to use MPI directly. 
Thus, {\em dart\_group\_union (group1, group2, newgroup)} is designed
to merge-sort the two input groups --- \textit{group1} and \textit{group2} automatically. In addition, inside the {\em dart\_group\_addmember (group1, unitid)},
we first perform {\em MPI\_Group\_incl (MPI$\_$COMM$\_$WORLD, 1, ranks, group2)},
where the array $ranks$ only consists of a single rank with an absolute \textit{unitID} that is expected to be added into the \textit{group2},
then followed by 
{\em dart\_group\_union (group1\_cpy, group2, group1)}.
Therefore, DART groups are guaranteed to be ordered once created. Using this method, 
not only can the specified unit be incorporated into the given DART group correctly,
but the semantics of DART groups also not get violated by the disordered characteristics of MPI group operations.
%TODO: The above is a little messy. I trid to clean. Someone please review. I think We are missing "ordered creation". (Colin)

\subsubsection{Team Translation}

Team is one of the central concepts in DART. The DART team plays a
similar role as the MPI communicator. From the unit point of view, a
team can be determined uniquely by \textit{teamID}, and for simplicity
the \textit{teamID} is not reused even after a team has been destroyed. 
%We use a hash table, called \textit{teams}, 
We use a linear array, called \textit{teams},
to record the one-to-one relationship between teams and their related communicators.
$teams[teamID]$ (a team specified by \textit{teamID}) is expected to store its corresponding communicator. 
However, it should be noted that the \textit{teamID} may become extremely large. 
Hence, the array --- \textit{teams} has to be large enough to meet the demand for gradually increasing \textit{teamID}.
It would be inefficient for DART to maintain such a large array when \textit{teamID} is used as an index of 
the array, because the teams can be destroyed during a program and therefore the space for the idle elements (corresponding to the destroyed teams)
in the \textit{teams} array can not be reused again.

% \todo{This is an implementation detail and not really necessary.}
To mitigate the aforementioned potential problem, 
we optimized the solution by introducing another array \textit{teamlist} with limited size, in which every element has the chance of indicating 
an existing team. 
%we optimized the solution by reusing the spaces of destroyed teams in the array \textit{teams}. 
%Thus, the size of the array can be limited. 
%We introduce another array \textit{teamlist} to mark the status of a team, with each element indicating whether an existing team is still in use. 
When creating a new team (e.g., team $a$), \textit{teamlist} is scanned linearly from the first element till the i-th element, 
where $teamlist[i] = -1$ indicating this is an empty slot. 
The slot is then allocated to the new team $a$ and initialized with the ID of team $a$. 
%In case that $teamlist[i]$ has the ID of a certain existing team, it is forbidden to rewrite this element. %TODO: I am not sure what the previous sentence is supposed to say (Colin) 
When team $a$ is destroyed, $teamlist[i]$ is reset back to $-1$.
%which means that the i-th element in both \textit{teamlist} and \textit{teams} can be reused.
The position of the given \textit{teamID} in \textit{teamlist} can be seen as a perfect index,
not only to locate the correct communicator in \textit{teams} but also for collective global memory pool and translation table. 
A detailed description of the latter will be given in the following section.

%% less specific formulation of above
%To mitigate the aforementioned potential problem, we optimized the solution by introducing another array \textit{teamlist} with limited size,
%in which every element has the chance of indicating an existing team.
%The position of the given team in \textit{teamlist} can be seen as a perfect index, not only to locate the correct communicator in \textit{teams}
%but also for collective global memory pool and translation table. A detailed description of the latter will be given in the following sections.

\subsubsection{Global Memory Management}
In DART, we have collective and non-collective global memory allocation.

DART non-collective global memory allocation is a local operation
which asks the  calling unit to allocate a block of globally accessible memory with given size.
However, MPI windows are created collectively across the corresponding communicator 
and therefore no one-to-one relationship between DART \textit{non-collective global pointers} and window objects exists. 
Hence,  
all the global memory blocks that are allocated with the DART non-collective allocation call have to be placed within a single pre-defined global window.

\begin{figure}[!ht]
\begin{center}
\includegraphics[scale=0.55]{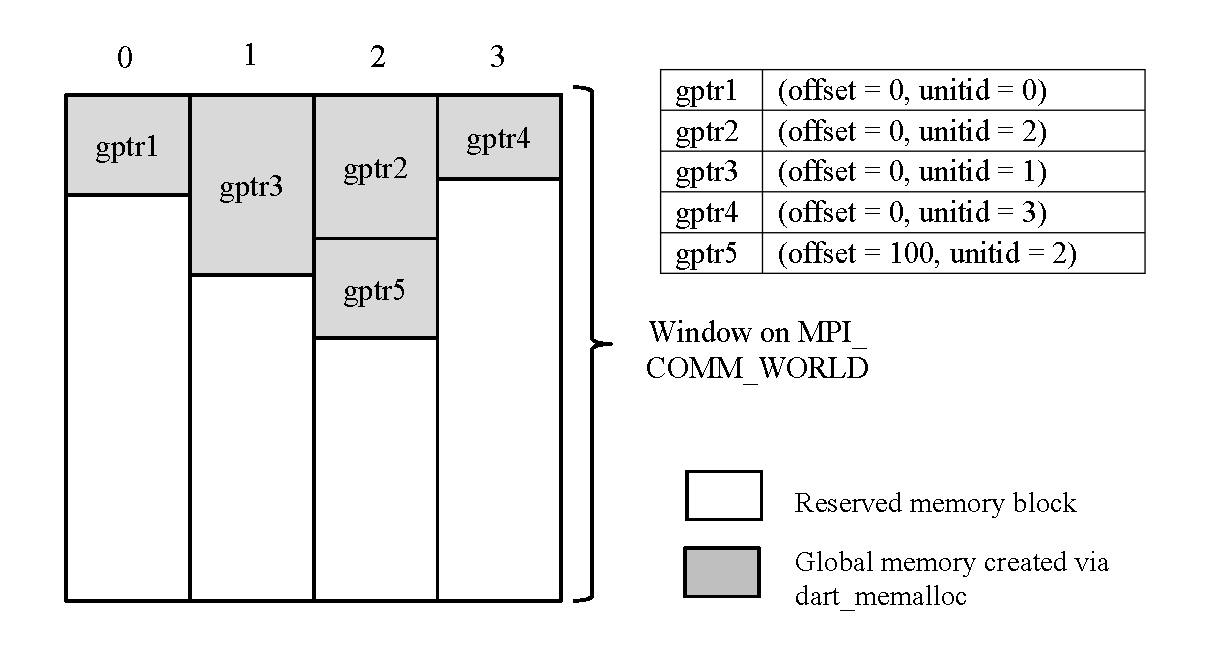}
\caption{Schematic Example of DART Non-Collective Global Memory Allocation}
\label{allo}
\end{center}
\end{figure}

As a result, for DART non-collective global memory allocation, we first reserve a memory block of sufficient size across all the running units. 
A global window is then created on {\em
MPI\_COMM\_WORLD}.
Finally, a call to DART non-collective global memory allocation starts a shared access epoch in the window for all participating units. 
Each unit manages its own partition of memory separately.
The offset in the \textit{non-collective global pointer} represents the displacement relative to the base address. 
Fig. \ref {allo} shows the method of handling DART non-collective global memory allocation.

\begin{figure}[h!]
\begin{center}
\includegraphics[scale=0.55]{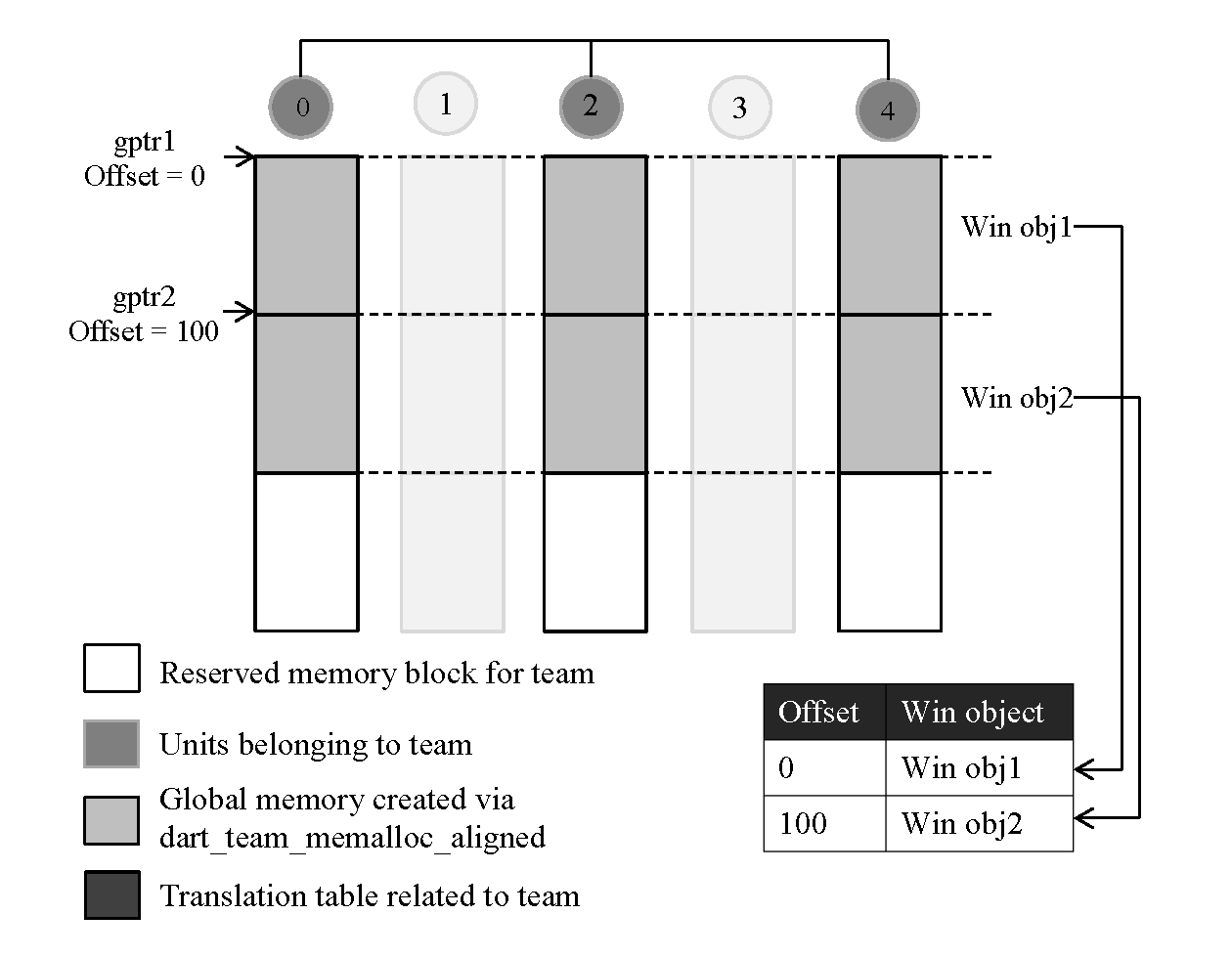}
\caption{Schematic Example of DART Collective Global Memory Allocation}
\label{collective}
\end{center}
\end{figure}

For DART collective global memory allocations, the offset to the corresponding window object is recorded in a translation table. 
Every team, upon creation, allocates an empty translation table 
and reserves a collective global memory pool for future DART collective global memory allocations.
The latter guarantees the possibility  of aligned allocations, leading to the identical offset for all units. 
As shown in Fig. \ref {collective}, an MPI window of requested size is created  every time a collective allocation is performed, 
and therewith a \textit{collective global pointer} is generated,
and finally the window object and offset are entered into the translation table. 
It is important to note that the offset in the returned \textit{collective global pointer}
represents the displacement relative to the base address of the memory 
region reserved for this team rather than the beginning of the sub-memory spanned by certain DART collective allocation. 

\subsubsection{Global Pointer Dereference and Unit Translation}

To use MPI as an underlying communication substrate, the location of
data and the ranks of the target processes needs to be
known. 

The location is given by DART global pointers, which are determined from the target unit, segmentation ID (equivalent to \textit{teamID}),
a specific offset and flags,
where the target unit is represented with an absolute \textit{unitID}.
In addition,
the determination of target rank depends on the type of DART global memory allocation: collective or non-collective,
which is identified according to the value of flags.

For MPI, the relative target ranks in the given communicators are entailed for launching the RMA operations,
Therefore, In the case that we refer to the \textit{collective global pointers},
we must translate the DART absolute \textit{unitIDs}
to the relative \textit{unitIDs} (ranks) in the given teams (communicators) for locating the correct target data.

From the section above, we can learn that the \textit{non-collective global pointers} are only active within
a pre-defined window associated with {\em MPI\_COMM\_WORLD}. Therefore,  
unlike the \textit{collective global pointers}, the \textit{non-collective global pointers}
can be trivially dereferenced without the unit translations. 

%since the MPI rank within the corresponding communicator can be trivially determined 
%from the DART absolute \textit{unitID} and \textit{teamID}.

 %The DART one-sided communications are performed based on DART global pointers, where the unit is represented by an absolute \textit{unitID} and the address to data in global memory is determined via offsets. 
%The offsets depend on the type of DART memory allocation: collective or non-collective. 

%Flags function like a signal, with 0 indicating DART non-collective global memory allocation$\/$free --- {\fontfamily{lmtt}\selectfont dart$\_$memalloc$\/$free} and 1 indicating DART collective global memory allocation$\/$free --- {\fontfamily{lmtt}\selectfont dart$\_$team$\_$memalloc$\_$aligned/dart$\_$team$\_$memfree}. 

%DART non-collective global memory allocation is a local operation
%which asks the  calling unit to allocate a block of globally accessible memory with given size.
%However, MPI windows are created collectively across the corresponding communicator 
%and therefore no one-to-one relationship between DART non-collective global pointers and window objects exists. 
%Hence, all the global memory blocks that are allocated with the DART non-collective allocation call have to be placed within a single pre-defined global window,
%and then the respective offsets are returned in the global pointers. 

%For DART collective global memory allocations, the offset to the corresponding window object is recorded in a  translation table. 
\subsubsection{One-sided and Collective Communication}

MPI RMA consists of RMA communication and synchronization calls. RMA communication
calls associated with {\em window} must proceed only within an
access epoch for this {\em window}, %RMA communication calls with argument {\em win} must occur within an access epoch for {\em win}, 
as depicted in Fig. \ref {sync}. However, the DART specification does not provide any concept for RMA synchronization. 
Therefore we need to start a shared access epoch on a given window before calling DART RMA communications, 
which is done automatically within DART collective global memory allocation and DART initialization calls.

MPI-3 extends the RMA communication interfaces with Request-based RMA communication, 
which associates a request handle with the RMA operations and enables us to test or wait for the completion of these requests using the functions --- {\em MPI\_Wait/Test/Waitall/Testall}. 
DART one-sided interfaces firstly perform the global pointer dereference, 
then do the unit translation only if the \textit{collective global pointers} are accessed, 
and finally execute MPI Request-based RMA operations. 
There is an MPI restriction implying that such operations are only valid within the MPI passive target epoch, 
which however does not impose any extra limitation on DART semantics.
As mentioned in a previous section, 
DART adopts the MPI-3 RMA passive target mode rather than the active mode.

The semantics of DART collective routines are the same as that of MPI.
Therefore, we can implement the DART collective interfaces straightforwardly by using the MPI-3 collective counterparts. 
Before calling the MPI-3 collective counterparts, we need to determine the communicator based on the given \textit{teamID}.

\subsubsection{Synchronization}
We implement the DART synchronization API using the MPI-3 RMA atomic memory access functionalities, based on a queuing mutex algorithm proposed by Mellor-Crummey and Scott \cite{Mellor}, 
which is proved to be a suitable one-sided mutual exclusion algorithm. 
This algorithm can be understood as a mechanism, namely list-based queuing lock (MCS lock).
In order to ensure the correctness of this mechanism, the atomicity of accessing to the mutexes has to be guaranteed accordingly.
It requires an atomic {\em fetch\_and\_op} (store/read) instruction and an atomic operation of {\em compare\_and\_swap}, 
which are provided by MPI-3. 
As stated earlier, the MPI-3 RMA shared access epoch mode is preferred to exclusive mode in the hope of enhancing efficiency.
Moreover, the characteristics of atomicity protect DART from conflicting accesses to memory. 

In DART, the lock creation operation is performed as a collective operation on a given team, and there can be multiple locks per team. %Locks are stored in  {\em dart\_lock\_struct}, which includes information on corresponding window objects, as well as communicator
Every unit using the locks allocates a compound record
containing a distributed queue --- \textit{list}, a \textit{non-collective global pointer} --- \textit{tail}, a \textit{teamID}, a window object, in which the queue chains DART units holding or waiting for this lock together.
%Every unit waiting to acquire this lock is written into a distributed queue --- list, a tail pointer, a \textit{teamID}, a window object and a Boolean flag. 
In practice, the queue functions like a global pointer to a shared variable stored with the next unit waiting on this queue for acquiring the lock, which guarantees FIFO ordering of lock acquisition. 

\begin{figure}[h!]
\begin{center}
\includegraphics[scale=0.55]{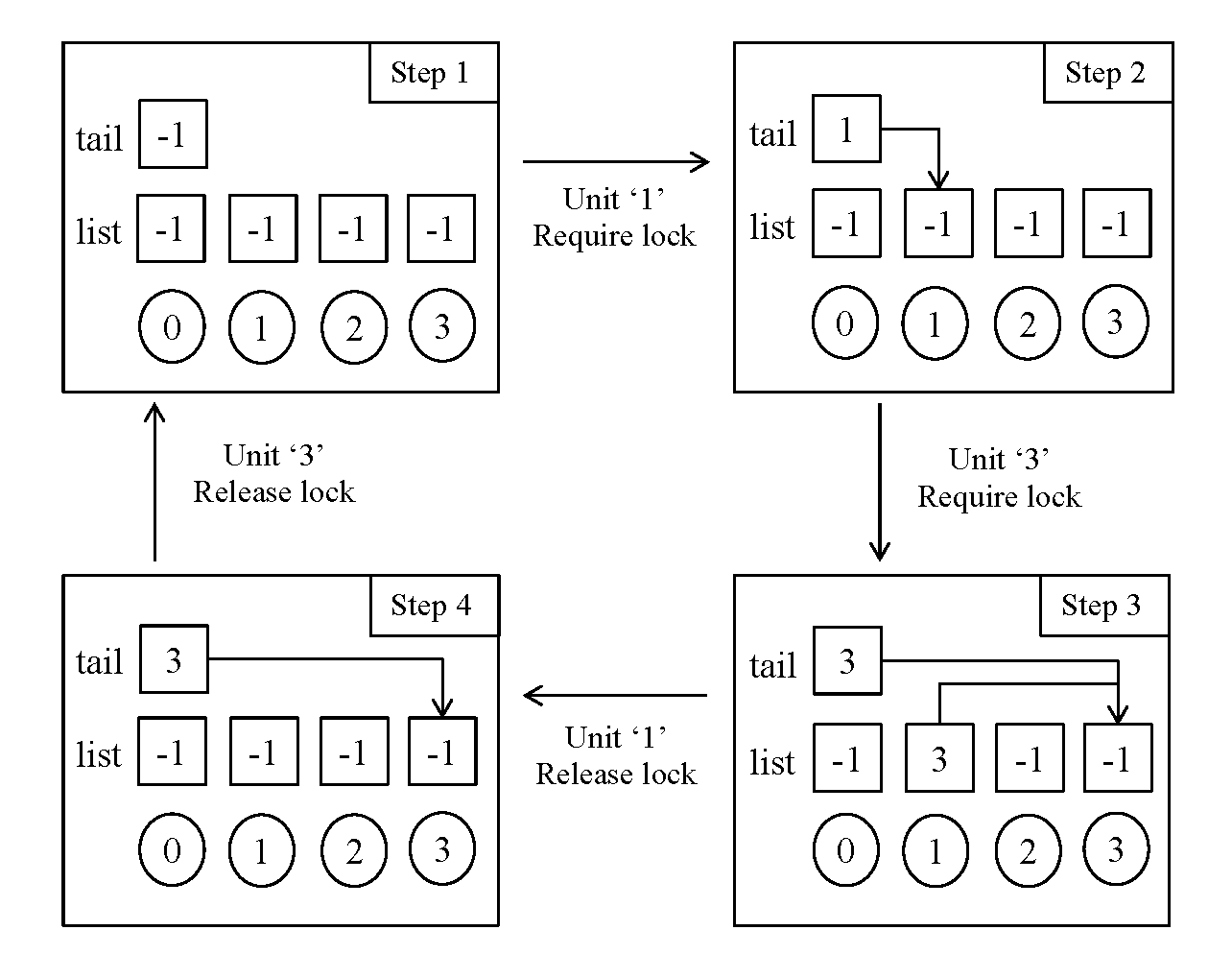}
\caption{Schematic Example of DART Synchronization Events}
\label{mutex}
\end{center}
\end{figure}

We create a block of global memory to store the \textit{tail} of the queue on the first unit, i.e., unit $0$ of team $a$ during the lock initialization via {\em dart\_memalloc}. We then allocate a block of global memory on all units of team $a$ along with the associated window object via {\em dart\_team\_memalloc\_aligned}. Each partition of the collective global memory (the DART distributed queue) is locally used to hold the next unit in the queue waiting for the lock. %As shown in Figure \ref{mutex}
Fig. \ref{mutex}, Step $1$ through Step $4$, illustrates the DART Lock/Unlock protocols using \textit{list} and \textit{tail}. Initially both \textit{tail} and \textit{list} point to $-1$, which means the lock is available and the waiting queue is still empty.

A lock acquisition is performed by unit $i$ via {\em dart\_lock\_acquire}. The atomic operation of {\em fetch\_and\_store} is applied, which consists of a series of actions. It first checks whether the lock has been acquired through referencing the \textit{tail}. In case that the lock is available, it acquires the lock and points the \textit{tail} to unit $i$. Otherwise,  it puts unit $i$ into the waiting queue. Once queued, the unit $i$ waits on an {\em MPI\_Recv} operation from its predecessor in the waiting queue.

Unlocking is performed via {\em dart\_lock\_release}. This function calls {\em compare\_and\_swap} to check whether the calling unit $i$ is the only unit in the queue. If this is the case, the \textit{tail} is set to $-1$. %The above operations are done as a single atomic action. 
Otherwise,  unit $i$ sends a zero-size notification to its successor in the waiting queue to announce the release of the lock. 

\section {Performance Evaluation}
\label {evaluation}

In the following section we evaluate the performance of DART-MPI. We start by defining the metrics on the basis of which we 
have evaluated the performance of the  runtime and then we describe the benchmark environment which covers the explanation of 
architecture and software environment of the machine on which the measurements are taken, and finally we present and interpret 
the performance results.
 
\subsection{Evaluation Metrics}
To assess the efficiency of the MPI based implementation of DART, we
have measured the \textit{Data Transfer Completion Time (DTCT)} and the \textit{Data Transfer Initiation Time (DTIT)} for 
blocking and non-blocking put/get calls respectively. We have also determined 
the bandwidth of the blocking and non-blocking variants of put and get operations provided by
DART. Here, we are mainly interested in quantifying the overheads with
respect to semantically equivalent operations done in pure MPI,
i.e. without the additional code due to DART. Thus, we vary only two
parameters, the message size, and the relative location of
communication partners. For the latter case we benchmarked the
following configurations:
\begin{itemize}
  \item {\em Intra-NUMA Performance:} The two processing units (PUs) are allocated on the same NUMA domain.
  \item {\em Inter-NUMA Performance:} The two PUs are allocated on distinct NUMA domains on the same node.
  \item {\em Inter-Node Performance:} The two PUs are allocated on distinct nodes.
\end{itemize}
Furthermore, we have used core pinning (i.e. each PU is pinned to a particular physical core) and strict memory containment per NUMA domain 
(i.e. a PU can allocate memory only on the local memory module of its assigned NUMA domain). 

For non-blocking operations, we have only measured the time for \textit{data transfer initiation}, whereas for bandwidth, 
the time for \textit{data transfer completion} (of many overlapping non-blocking operations) is considered. 
The reason for only measuring the DTIT is that the non-blocking calls allow to hide the 
time of data transfer by overlapping it with some computation, because 
these calls return immediately after initiating the transfer. We are not interested in the time spent after the transfer initiation till 
its completion. Whereas for bandwidth measurements, we want to make sure that the data is actually transferred from source to destination, 
not on the basis of only transfer in progress.

\subsection{Benchmark environment}

\begin{figure}
\includegraphics[width=8cm]{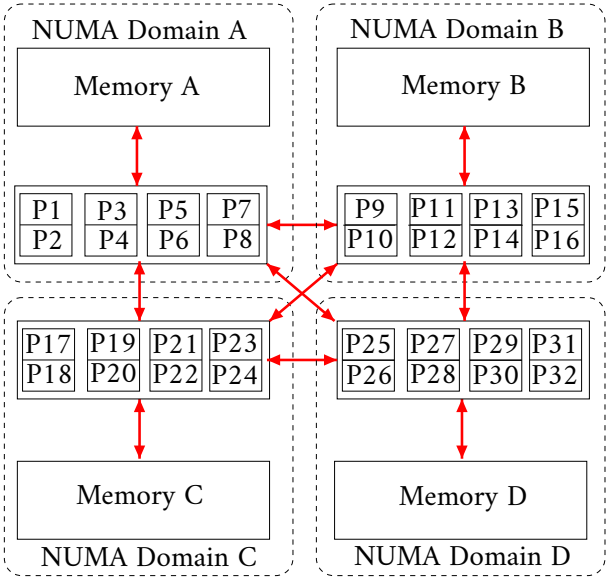}
\caption{Topology of a Single Compute Node of the Hermit System.}
\label{IL_topology}
\end{figure}

The presented benchmarks have been produced on Hermit, a Cray XE6 system at HLRS.  
Each node of Hermit features two AMD Opteron 6276 (Interlagos) processors, which are clocked at 2.3GHz.
An interlagos processor is composed of two orchi dies (each consists of 4 Bulldozer modules - 2 cores per module), such that each processor has 16 cores which are divided 
into two NUMA domains. Therefore, each node of the system is comprised of 32 cores (4 NUMA domains - 8 cores per NUMA domain). The nodes are inter-connected using 
Cray's high speed network 'Gemini'. The topology of a single node is shown in Fig.~\ref{IL_topology}.

The code was built using the Cray compiler (version 8.2.5 - utilizing Cray MPICH2 MPT 6.2.1). 
For inter-NUMA benchmarks, NUMA domains on different processors are selected.
 
%When running simulations on partially filled compute nodes, i.e. the
%number of processing elements (PE) is smaller than the number of
%available CPUs, distributing PEs evenly among the numa
%nodes increases the bandwidth per PE. Furthermore, pinning PEs to a given CPU or
%restricting them to a single numa node and allowing memory allocations only
%within that particular node prevents memory accesses across numa regions.

For a full documentation and technical
specification of the hardware platform, the reader is referred to the HLRS's wiki page \cite{Hermit}
%\footnote{https://wickie.hlrs.de/platforms/index.php/Cray\_XE6}.

\subsection{Results}
As mentioned, we have measured the DTCT and DTIT for blocking and
non-blocking operations (put and get), respectively, and compared those
to a pure MPI implementation. Similarly the bandwidth of these operations is
computed and examined in contrast with the corresponding MPI implementation.
All benchmarks are averaged over multiple
executions and the measurement errors are estimated from the statistical
standard deviation. 
%All measurements shown are thus averages over many runs. 
The standard deviation in general is small, typically less than
$10\%$ on data points. We do not show error bars in the figures in
order to keep them legible. We also did several sets of measurements
on different days. However, we present only one such set as the others
show consistent results. In order to quantify the overheads rigorously, the data is fitted to different models. In
particular, here we quote numbers from a model that assumes a constant
overhead between MPI and DART, i.e. $t_{DART}(m) - t_{MPI}(m) = f(m) =
c$, with $m$ as message size. We have also tested with models that allow
the overhead to vary with message size, but found consistent results.

\begin{figure}[h!]
\centering
\includegraphics[width=3in]{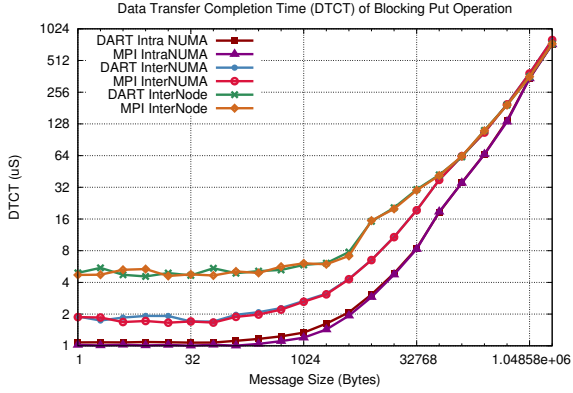}
\caption{Comparison of the DTCT of the Blocking Put Operation}
\label{b_put_latency}      
\end{figure}

\begin{figure}[h!]
\centering
\includegraphics[width=3in]{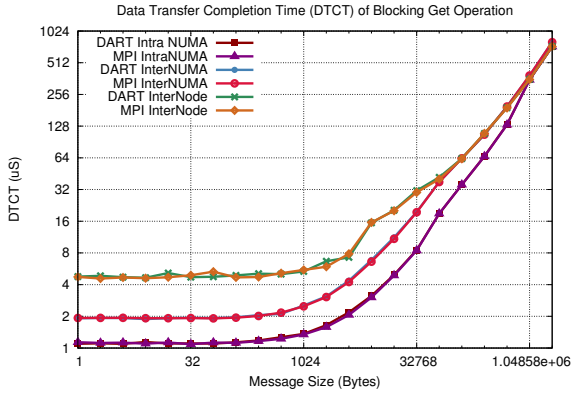}
\caption{Comparison of the DTCT of the Blocking Get Operation}
\label{b_get_latency}      
\end{figure}

Figures \ref{b_put_latency} and \ref{b_get_latency} show the DTCT of
blocking put and get operations of DART and native MPI,
respectively. We have varied the message size  from 
1 to 2\textsuperscript{21} bytes and repeated measurements for all
three different cases of relative process placement.  Just by looking
at the figures, one can see that the overhead of DART is very small compared 
to pure MPI.  The analysis of the model indeed shows that, given the
measurement error, all data is consistent with vanishing
overheads. Only in the case of inter-NUMA put operations could we
measure a statistically significant overhead of $(81 \pm 6)\;
ns$ across all messages sizes. However, this is equal to a small
fraction of the DTCT, which is in the order of $1\;\mu s$.   
%It can be seen that the DART shows almost the same performance as the native MPI. 
%The overhead caused by the additional code in DART is negligible in all cases 
%(intra-NUMA, interNUMA and inter-Node).

Notably, the Cray-MPI messaging protocol changes from \textit{eager E0} (i.e. no copying 
of data to buffer) to \textit{eager E1} (i.e. data is copied into internal MPI buffers 
on both the send and receive side) when the message size is greater than 4KB. 
The impact due to this change in messaging protocol is visible in the figures 
\ref{b_put_latency} and \ref{b_get_latency}, where there is a sudden jump in the 
DTCTs of operations between 4KB and 8KB.

\begin{figure}[h!]
\centering
\includegraphics[width=3in]{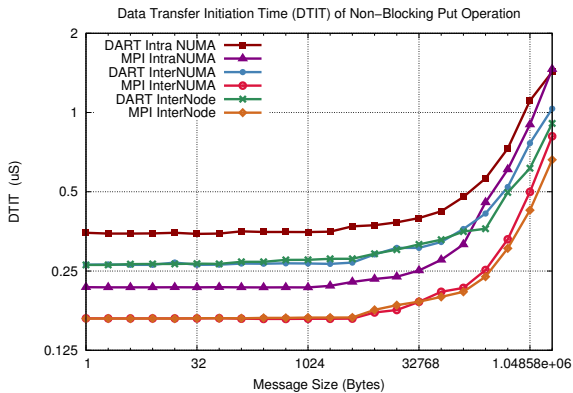}
\caption{Comparison of the DTIT of Non-blocking Put Operation}
\label{nb_put_latency}      
\end{figure}

\begin{figure}[h!]
\centering
\includegraphics[width=3in]{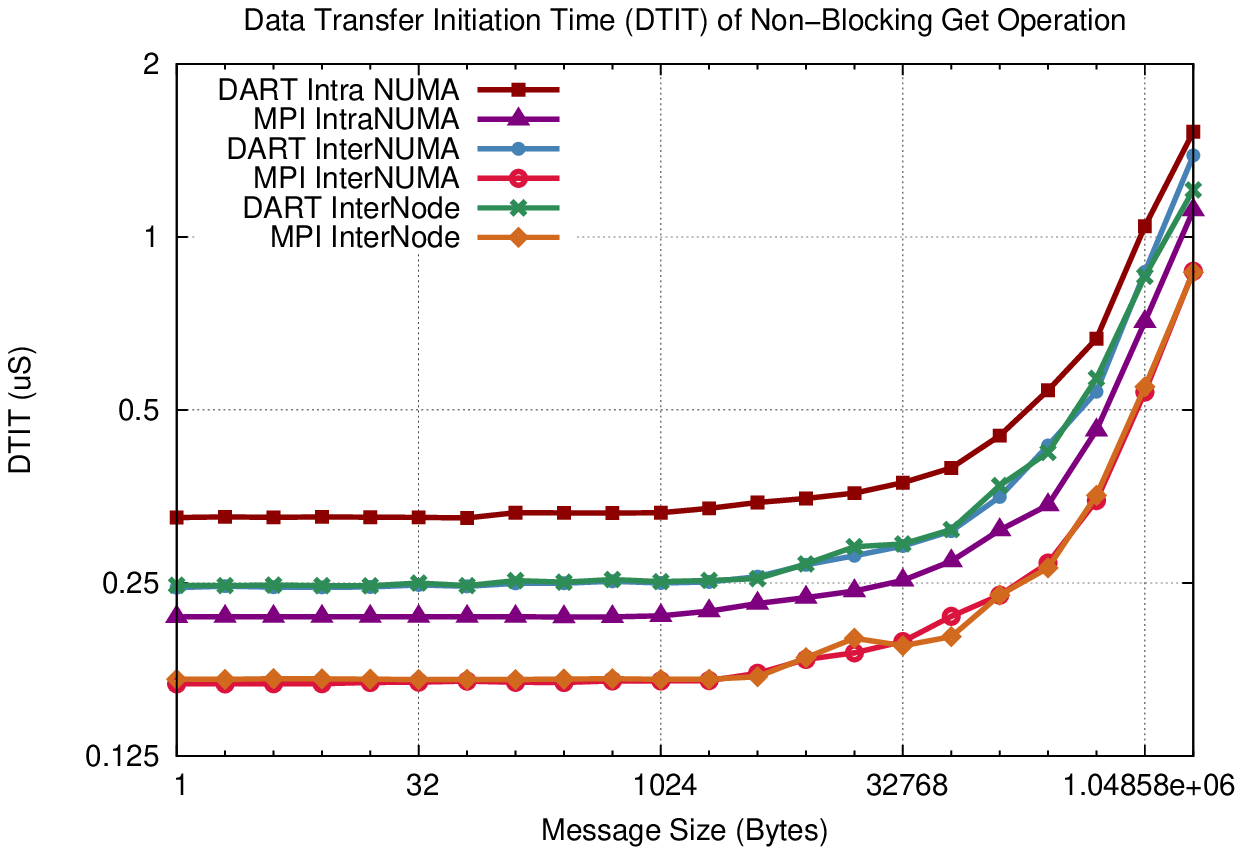}
\caption{Comparison of the DTIT of Non-blocking Get Operation}
\label{nb_get_latency}      
\end{figure}

Next, we turn our attention to non-blocking operations. 
The DTITs of the non-blocking put and get operations of DART and
native MPI are shown in figures \ref{nb_put_latency} and
\ref{nb_get_latency} respectively. The overhead for non-blocking put is around
$100\;ns$ with a standard deviation of a few percent only. Inside the same
NUMA domain, our models show a slightly larger overhead of
$130\;ns$. Similarly, non-blocking get operations have an overhead of
around $80\;ns$ in general, with a slightly larger value of $110\;ns$
when communicating inside the same NUMA domain.

%DART utilizez the MPI_Rput and MPI_Rget for the 
%implementation of put and get operations respectively, the use of MPI_R* variants infuses an overhead of approx. 100nS compared to their conterparts variants which do not use a 
%request handle to track a particular RMA operation (as shown in figures \ref{nb_put2} and \ref{nb_get2} respectively). 

\begin{figure}[h!]
\centering
\includegraphics[width=3in]{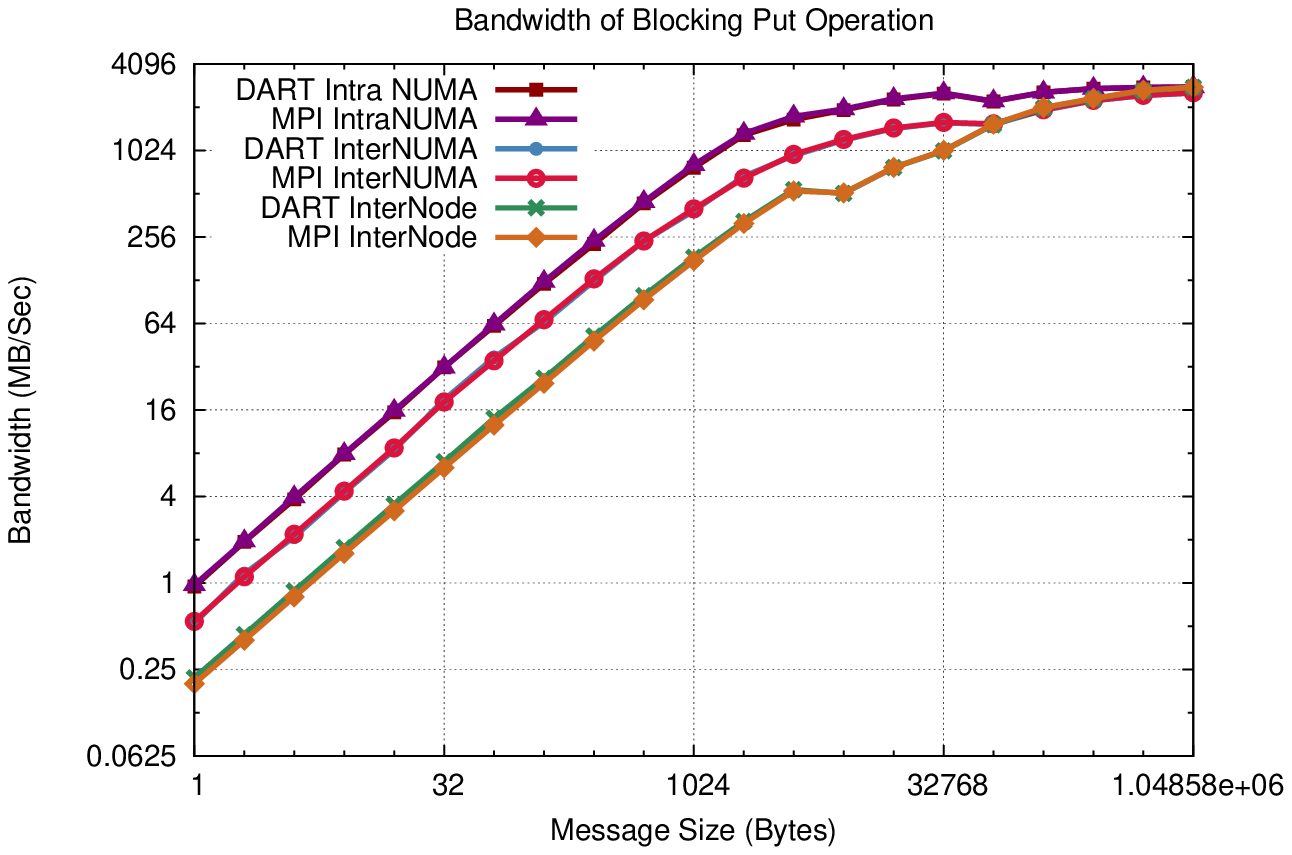}
\caption{Comparison of the Bandwidth of the Blocking Put Operation}
\label{b_put_bw}      
\end{figure}

\begin{figure}[h!]
\centering
\includegraphics[width=3in]{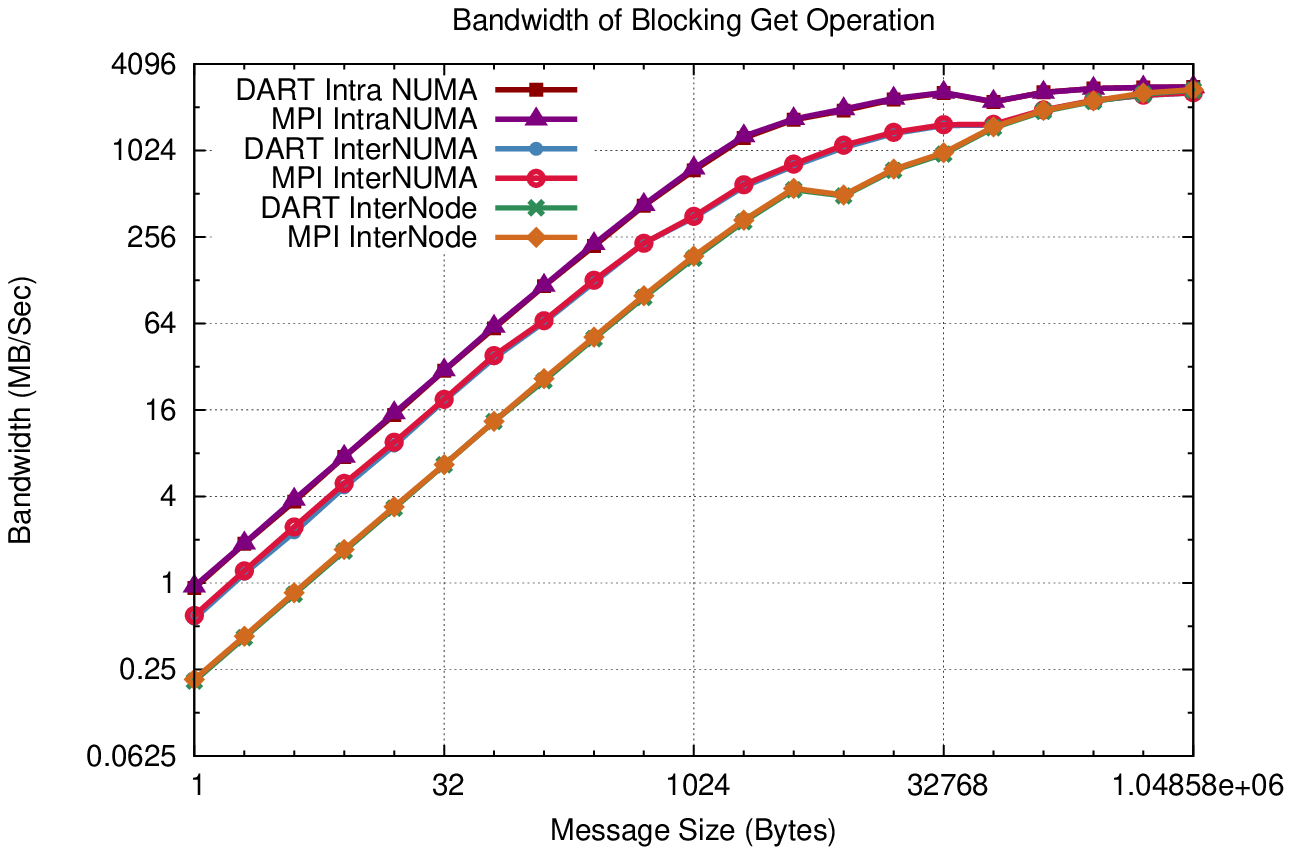}
\caption{Comparison of the Bandwidth of the Blocking Get Operation}
\label{b_get_bw}      
\end{figure}

\begin{figure}[h!]
\centering
\includegraphics[width=3in]{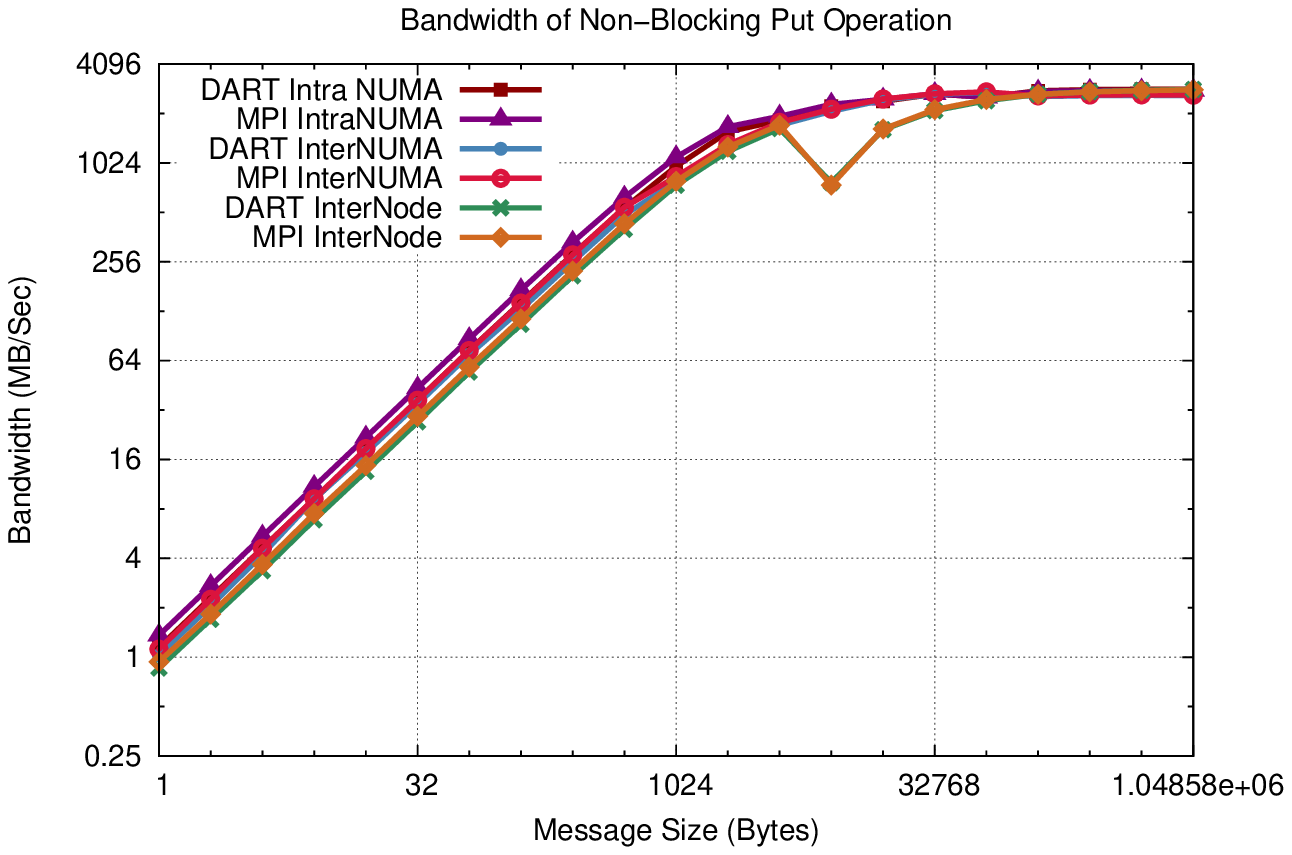}
\caption{Comparison of the Bandwidth of the Non-blocking Put Operation}
\label{nb_put_bw}      
\end{figure}

\begin{figure}[h!]
\centering
\includegraphics[width=3in]{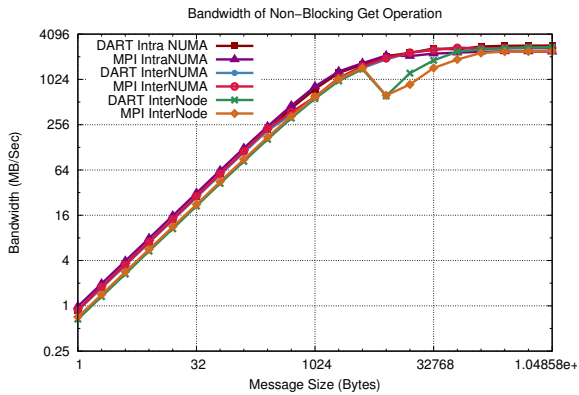}
\caption{Comparison of the Bandwidth of the Non-blocking Get Operation}
\label{nb_get_bw}      
\end{figure}

The results of bandwidth benchmarks for the various RMA operations are
shown in figures \ref{b_put_bw}--\ref{nb_get_bw}, respectively. As
expected from the previous analysis, the performance of DART is
comparable to pure MPI as overheads are negligible in most cases. In
fact, the choice of protocol used in the Cray MPI implementation,
i.e. E0 versus E1, seems to have a larger impact on the bandwidth (as
seen from the sudden drop in bandwidth, e.g. Fig. \ref{nb_get_bw}
around $8KB$) than the difference between DART and MPI.

\section {Conclusions and Further Work}
We have presented a preliminary implementation of DART with MPI-3 as its lower-layer communication system. Although with the improvement
and extension in MPI-3 RMA, there are still some mismatches between DART and MPI in the semantics, e.g., 
DART team versus MPI communicator and 
DART global pointer versus MPI window object, which have to be resolved. 

In addition, as we have seen from the results, DART has approximately the same performance as MPI for blocking operations. For non-blocking operations, the overhead is statistically 
significant and around in the order of $100\;ns$. 
This overhead is prominent for small messages, up to 128KB it is around one third of the total time taken by the DART operation. As the overhead is constant, the impact lessens with 
growing message size. 
%, but negligible for large message sizes.
%Thus, we can conclude that the extra overhead will not impair the overall performance of DART in latency and bandwidth to a great extent. 

In the future, we plan to enable the MPI-3 shared-memory
window option for DART, which provides true zero-copy mechanisms, as opposed to traditional single-copy mechanisms.
An early implementation using MPI-3 shared memory window shows promising preliminary results:
especially for small message sizes, intra- and inter-NUMA communication becomes a lot more efficient. We are currently 
performing a detailed analysis in order to guarantee the quality and correctness of this implementation.
There are potential scalability issues existing in DART. For instance, DART currently map a teamID to an entry in the \textit{teamlist}
through linearly scanning this \textit{teamlist}, in which case the overhead brought by the scanning 
can be significant when the teamlist is extremely large. However, linked list can be a straightforward alternative for 
teamlist. In addition, We always allocate a global memory block used as \textit{tail} 
on the unit $0$ in a certain team every time when a lock on this team is initialized,
which will lead to a communication congestion on the unit $0$ when multiple separate locks are allocated within this team. 
We intend to balance the distribution of the \textit{tail} between all participating units of a team.

%As we have shown in this paper, DART already is very competitive and we will improve it even further.  
%Thus, we believe DASH is becoming a highly competitive programming model: providing and exploiting 
%advanced PGAS features, with only marginal communication overheads compared to MPI.

\label {conclusion}

\section*{Acknowledgment}
The authors would like to thank Andreas Kn\"upfer, Denis H\"unich, and
Andr\'e Gr\"otzsch for fruitful discussion on the DASH runtime
design. We gratefully acknowledge funding by the German Research
Foundation (DFG) through the German Priority Programme 1648 Software
for Exascale Computing (SPPEXA).
%  under grant agreements XXX. 

\bibliographystyle{plain}
\bibliography{paper}

\end{document}